\documentclass{PoS}
\usepackage[utf8]{inputenc}
\usepackage{amsmath}
\usepackage{amsfonts}
\usepackage{amssymb}
\usepackage{graphicx}
\usepackage{epsfig}
\usepackage{tabularx}
\usepackage{lineno}
\usepackage[section]{placeins}

\title{Studying Neutral Current Elastic Scattering and the Strange Axial Form
             Factor in MicroBooNE}

\ShortTitle{NC Elastic Scattering in MicroBooNE}

\author{\speaker{Katherine Woodruff for the MicroBooNE Collaboration}\\
        New Mexico State University\\
        E-mail: \email{kwoodruf@nmsu.edu}}

\abstract{One of the least constrained contributions to the neutral current
(NC) elastic neutrino-proton cross section is the strange axial form factor,
which represents the strange quark spin contribution to the spin structure of
the proton. This becomes the net strange spin contribution, $\Delta s$, in the
limit when the negative four-momentum transfer squared ($Q^2$) is zero. The
strange axial form factor can be determined by studying NC elastic scattering
events in the MicroBooNE detector.  MicroBooNE's unique ability to detect
low-energy protons is expected to allow the reconstruction of these events with
a $Q^2$ as low as 0.10 GeV$^2$ and to determine the strange axial form factor
in a model-independent approach. We present a selection of neutral current
elastic events in a subset of MicroBooNE neutrino data, as well as our plan to
extract the strange part of the axial form factor and $\Delta s$ from this
selection in the full data set.}

\FullConference{23rd International Spin Physics Symposium - SPIN2018 -\\
        10-14 September, 2018\\
        Ferrara, Italy}

\begin{document}

\section{Introduction}
\label{sec:intro}

The structure of a nucleon is more complicated than the three familiar up and
down valence quarks. These three quarks only account for a small percent of the
nucleon mass. The gluons that bind the quarks split into quark-antiquark pairs
of up, down, and strange flavor. The remainder of the nucleon mass is carried
by this quark-gluon sea. The structure of the sea and how its elements combine
with the valence quarks to give the nucleon its measured structure are not
precisely known.

The net spin of the proton comes from a combination of the spin and orbital
momentum of the quarks and gluons. The net contribution from the spin of
strange quarks and antiquarks, $\Delta s$, is defined as
\begin{equation*}
  \Delta s = \int_0^1 \Delta s(x) \, dx
\end{equation*}
\begin{equation*}
  \Delta s(x) = \sum_{r=\pm 1} r[s^{(r)}(x) + \bar{s}^{(r)}(x)] \,,
\end{equation*}
where $s(\bar{s})$ is the spin-dependent parton distribution function of the
strange (anti)quark, $r$ is the helicity of the quark relative to the proton
helicity and $x$ is the Bjorken scaling variable~\cite{Alberico01}.  In the
static quark model this value is zero.

In the 1980s the European Muon Collaboration~\cite{Ashman89} and several
subsequent experiments found that the Ellis-Jaffe Sum Rule was violated in
polarized, charged-lepton, inclusive, deep inelastic scattering (DIS). The
Ellis-Jaffe sum rule~\cite{Ellis74} assumes that SU(3) flavor symmetry is valid
and that $\Delta s = 0$. For the results to be consistent with exact SU(3)
flavor symmetry, $\Delta s$ must be \textit{negative}. Follow-up measurements
using charged-lepton semi-inclusive deep inelastic scattering have been
consistent with $\Delta s = 0$, but these determinations of $\Delta s$ are
highly dependent on the fragmentation functions used~\cite{Aidala12}.

An independent determination of $\Delta s$ can be made using neutral-current
(NC) elastic neutrino-proton scattering. The NC elastic cross section depends
directly on $\Delta s$ and no assumptions about SU(3) flavor symmetry or
fragmentation functions are needed.

\subsection{Neutral Current Elastic Neutrino-Proton Scattering ($\nu p \rightarrow \nu p$)}
To see how the dependence on $\Delta s$ arises, it is illustrative to look at
the simplest case of free nucleon scattering. In reality, the neutrinos are
interacting with nucleons in an argon nucleus which can modify the scattering
cross section and the observed final state. The neutral-current elastic (NCE)
neutrino-proton cross section for free nucleon scattering is given
by~\cite{Alberico01},
\begin{align*}
\begin{split}
\left(\frac{d\sigma}{dQ^2}\right)_{\nu}^{NC} &=\frac{G_F^2}{2\pi} \left[ \frac{1}{2}y^2(G_M^{NC})^2+\left(1-y-\frac{M}{2E}y \right) \frac{(G_E^{NC})^2+\frac{E}{2M}y(G_M^{NC})^2}{1+\frac{E}{2M}y} \right.\\
&+ \left. \left(\frac{1}{2}y^2 + 1 - y + \frac{M}{2E}y \right)(G_A^{NC})^2 \pm 2y \left(1-\frac{1}{2}y \right) G_M^{NC}G_A^{NC} \right]\,,
\end{split}
\end{align*}
where $G_A^{NC}$ is the neutral-current axial form factor, $G_E^{NC}$ is the
neutral-current electromagnetic form factor, and $G_M^{NC}$ is the
neutral-current magnetic form factor. These form factors represent the finite
structure of the proton. The axial form factor represents the spin structure of
the proton. Individual quark contributions to $G_E$ and $G_M$ have been
measured in electron-nucleon scattering, from which we can predict $G_M^{NC}$
and $G_E^{NC}$. The strange quark contribution to the electromagnetic form
factors has been determined to be small. This allows us to determine $G_A^{NC}$
from NCE interactions.

To determine the net strange quark spin contribution we can write $G_A^{NC}$ in
terms of quark flavor and extrapolate to $Q^2 = 0$
\begin{align*}
  G_A^{NC}(Q^2) &= \frac{1}{2}G_A^{CC}(Q^2) + \frac{1}{2}G_A^s(Q^2) \,, \\
  G_A^{NC}(Q^2=0) &= \frac{1}{2}(\Delta u - \Delta d) - \frac{1}{2}\Delta s \,,
\end{align*}
where $(\Delta u - \Delta d) = g_A$ is the weak coupling constant which has
been measured in neutron decay, and $G_A^{CC}$ is the charged-current axial
form factor which contains information about the up and down quark spin
distributions.

While nuclear effects can modify the final state, the signal of a NCE proton
interaction is ideally a single proton track. Without a vertex to help identify
the track as coming from a neutrino interaction, they are difficult to
reconstruct. Everything that we know about this interaction, including momentum
transfer, comes from the single proton track. Since we want to extrapolate
$G_A^{NC}(Q^2)$ to zero, we need to be able to reconstruct very low energy
protons. We estimate that we will be able to detect a proton track if it
traverses at least 2.5~cm ($\sim 8$ wires) in MicroBooNE. This is the range of
a 50~MeV kinetic energy proton in liquid argon, which corresponds to a NC
elastic interaction with $Q^2\sim 0.10\, $GeV$^2$.

\section{Model}
\label{sec:model}

  The model of neutrino interactions in the MicroBooNE detector, including the
  expected neutrino flux, the neutrino-nucleon cross section, the effect of the
  nucleon in an argon nucleus, and the detector response, is simulated in the
  MicroBooNE software. The initial neutrino-argon interactions are simulated
  using the GENIE neutrino generator~\cite{Andreopoulos09}, and the final state
  particles are propagated through the detector geometry using
  Geant4~\cite{Agostinelli02}. In the model, only the neutrino-nucleon cross
  section is affected by the strange axial form factor. This
  allows us to determine how the simulation would change due to a change in the
  neutrino-nucleon cross section by calculating the ratio of the new cross
  section to the cross section used in the simulation. This method is referred
  to as \textit{reweighting}.

  Even though the ``signal" that we are optimizing for is NC elastic
  neutrino-proton interactions in the TPC, other interactions are affected by
  the strange axial form factor and the form of the elastic cross section in
  general. The cross sections for background NC elastic neutrino-neutron
  interactions and any NC elastic interactions that occur outside of the TPC
  are affected by the new strange form factor and need to be calculated.
  Charged current background events aren't affected by the strange axial form
  factor, but we calculate new charged current quasi-elastic (CCQE) cross
  sections for these events so that the cross section parameterization and form
  factor models match the neutral current models that we are using.

  \subsection{Cross Section Model}
  \label{sec:xsecmodel}

  For reweighting both the NC elastic and CCQE cross sections, we use the
  Llewellyn-Smith form:
  \begin{equation}
    \frac{d\sigma}{dQ^2} = \frac{G_F^2}{2\pi E_{\nu}^2}\left[A \pm BW + CW^2 \right]
  \end{equation}
  where the $+(-)$ is for (anti)neutrino scattering, $G_F$ is the Fermi
  coupling constant, $E_{\nu}$ is the incoming neutrino energy, $W =
  \frac{4E_{\nu}}{M_p} - \tau$, $\tau = \frac{Q^2}{4M_p^2}$, and
  \begin{align}
    A &= \frac{(m_l^2 + Q^2)}{4}\left[G_A(1+\tau) - (F_1^2 - \tau F_2^2)(1-\tau) + 4\tau F_1 F_2 \right] \\
    B &= -\frac{Q^2}{4}G_A (F_1 + F_2) \\
    C &= \frac{Q^2}{64\tau}\left[G_A^2 + F_1^2 + \tau F_2^2 \right] \,.
  \end{align}
  Here $m_l$ is the outgoing lepton mass and $G_A$, $F_1$, and $F_2$ are the
  axial, Dirac, and Pauli form factors, respectively. The form factors are
  different for NC  and CC elastic scattering, with
  \begin{align}
    F_{1,2}^{NC} &= \frac{1}{2}F_{1,2}^{CC} \,, \\
    G_A^{NC} &= \frac{1}{2}G_A^{CC} + \frac{1}{2}G_A^s \,,
  \end{align}
  where $G_A^s$ is the strange quark contribution to the
  axial form factor.

  \subsection{Form Factor Model}
  \label{sec:ffmodel}

  For the form factor models, we use a z expansion parameterization for all
  three of the form factors described in~\cite{Ye2018}~and~\cite{Meyer2016}.
  These parameterizations are made by mapping the negative four-momentum
  transfer squared, $Q^2$, onto a domain where the strange axial form factor is
  analytic. This gives a new variable
  \begin{equation}
    z(Q^2,t_{cut},t_0) = \frac{\sqrt{t_{cut} + Q^2} - \sqrt{t_{cut} -
        t_0}}{\sqrt{t_{cut} + Q^2} + \sqrt{t_{cut} - t_0}} \,,
  \end{equation}
  where $t_{cut}$ is the first pole and $t_0 < t_{cut}$ is an arbitrary number
  which can be optimized for the $Q^2$ range of the data.  The new variable,
  $z$, is guaranteed to be small with the optimal choice of $t_0$, so a Taylor
  expansion around it should converge. The form factors are defined by the
  Taylor expansion as
  \begin{equation}
    G(Q^2) = \sum_{k=0}^{k_{max}} a_k z(Q^2)^k \,,
  \end{equation}
  where $a_k$ are the coefficients that are fit to data.
  
  The fits that we use for the Dirac and Fermi form factors are
  in~\cite{Ye2018}, and the fit for the charged-current axial form factor is
  in~\cite{Meyer2016}.

  We can use the same method to determine the strange part of the
  neutral-current axial form factor from the number of NC elastic events in
  MicroBooNE. We use $t_{cut} = 4m_{\pi}$, the four-pion threshold which is the
  first pole in the axial form factor. If we set $k_{max} = 2$, this becomes a
  simple three parameter fit to the data
  \begin{equation}
    G_A^s(Q^2) = a_0^s + a_1^s\, z(Q^2) + a_2^s\, z(Q^2)^2 \,.
  \end{equation}
  We can redefine the strange axial mass, $M_A^s$, in terms of the slope
  of $G_A^s(Q^2)$ at $Q^2 = 0$ as it is in the dipole form
  \begin{equation}
    M_A^s \equiv \sqrt{\frac{2G_A^s(0)}{(G_A^s)'(0)}} = \sqrt{\frac{2a_0^s}{a_1^s}} \,.
  \end{equation}
  Since we also know that $G_A^s(Q^2 = 0) \equiv \Delta s$, we can write the
  first two coefficients in terms of physical parameters
  \begin{equation}
    a_0^s = \Delta s, \hspace{5mm} a_1^s = \frac{2\Delta s}{(M_A^s)^2} \,.
  \end{equation}

\section{Proton track identification \label{sec:protonid}}

This section has been adapted from Ref.~\cite{UBprotonid}.

\subsubsection{Gradient decision tree boosting}
To identify proton tracks, we use a gradient-boosted decision tree classifier.
We chose to use decision trees because they are easily interpretable and the
inputs can be a mix of numeric and categorical variables. Below is a short
description of gradient tree boosting. A more detailed description can be found
in the documentation for the XGBoost\cite{Chen16} software library that was
used. 

A decision tree can be thought of as a series of if/else statements that
separate a data set into two or more classes. The goal of each cut is to
increase the information gain. For numerical variables any cut value can be
selected by the tree.  At each node of the tree, a split is chosen to maximize
information gain until a set level of separation is reached.  At the terminus
of the series of splits, called a leaf, a class is assigned.

Two weaknesses of decision trees are their tendency to over fit the training
data and the fact that the output is a class label and not a probability.
Gradient-boosting addresses both of these issues by combining many weak
classifiers into a strong one. Each weak classifier is built based on the error
of the previous one. For a given training set, whenever a sample is classified
incorrectly by a tree, that sample is given a higher importance when the next
tree is being created.  Mathematically, each tree is training on the gradient
of the loss function. After all of the trees have been created, each tree is
given a weight based on its ability to classify the training set, and the
output of the gradient-boosted decision tree classifier is the probability that
a sample is in a given class.

\subsubsection{The decision tree model}
We created a multi-class gradient-boosted decision tree classifier, using the
XGBoost software library, to separate five different track types: any proton
track, muons or pions from BNB neutrino interactions, tracks from
electromagnetic showers from BNB interactions, and any non-proton track
produced by a cosmic ray interaction. The classifier takes reconstructed track
features as input and outputs a probability of the track having been produced
by each of the given particle types. The reconstructed features are based on
the track's geometric, calorimetric, and optical properties from MicroBooNE TPC
and PMT systems.

The training data that we use to make the decision trees comes from Monte Carlo
simulation. The BNB interactions are simulated using the GENIE neutrino
generator, and cosmic interactions are simulated using the CORSIKA cosmic ray
generator~\cite{Heck98}. The particles generated by GENIE and CORSIKA are
passed to Geant4 where they are propagated through a simulated MicroBooNE
detector. For training and testing of the trees we only use tracks that were
reconstructed in LArSoft.

\begin{figure}[!htb]
  \centering
  \begin{minipage}[t]{16pc}
    \includegraphics[scale=0.34]{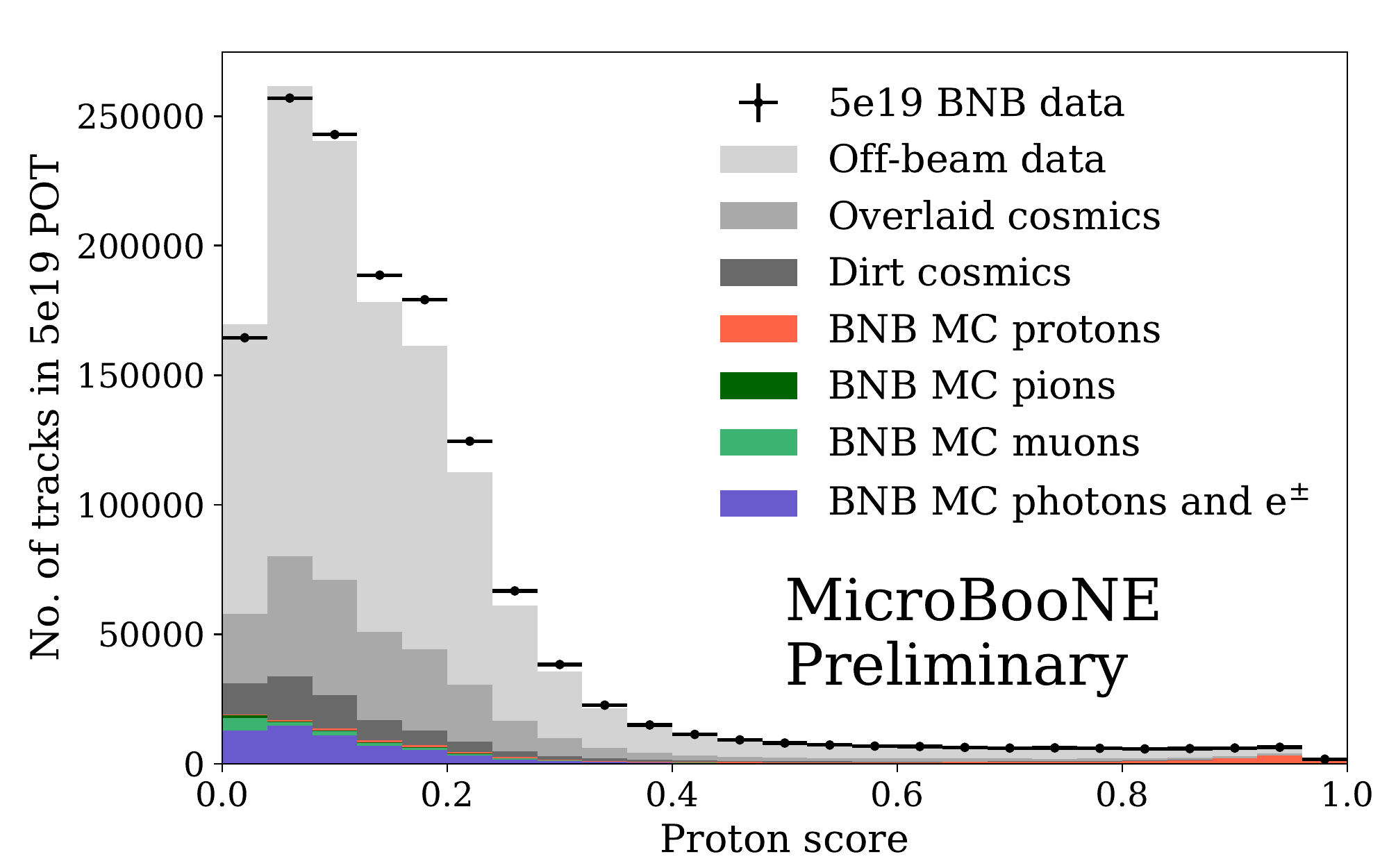}
  \end{minipage}\hspace{2pc}%
  \begin{minipage}[t]{16pc}
    \includegraphics[scale=0.34]{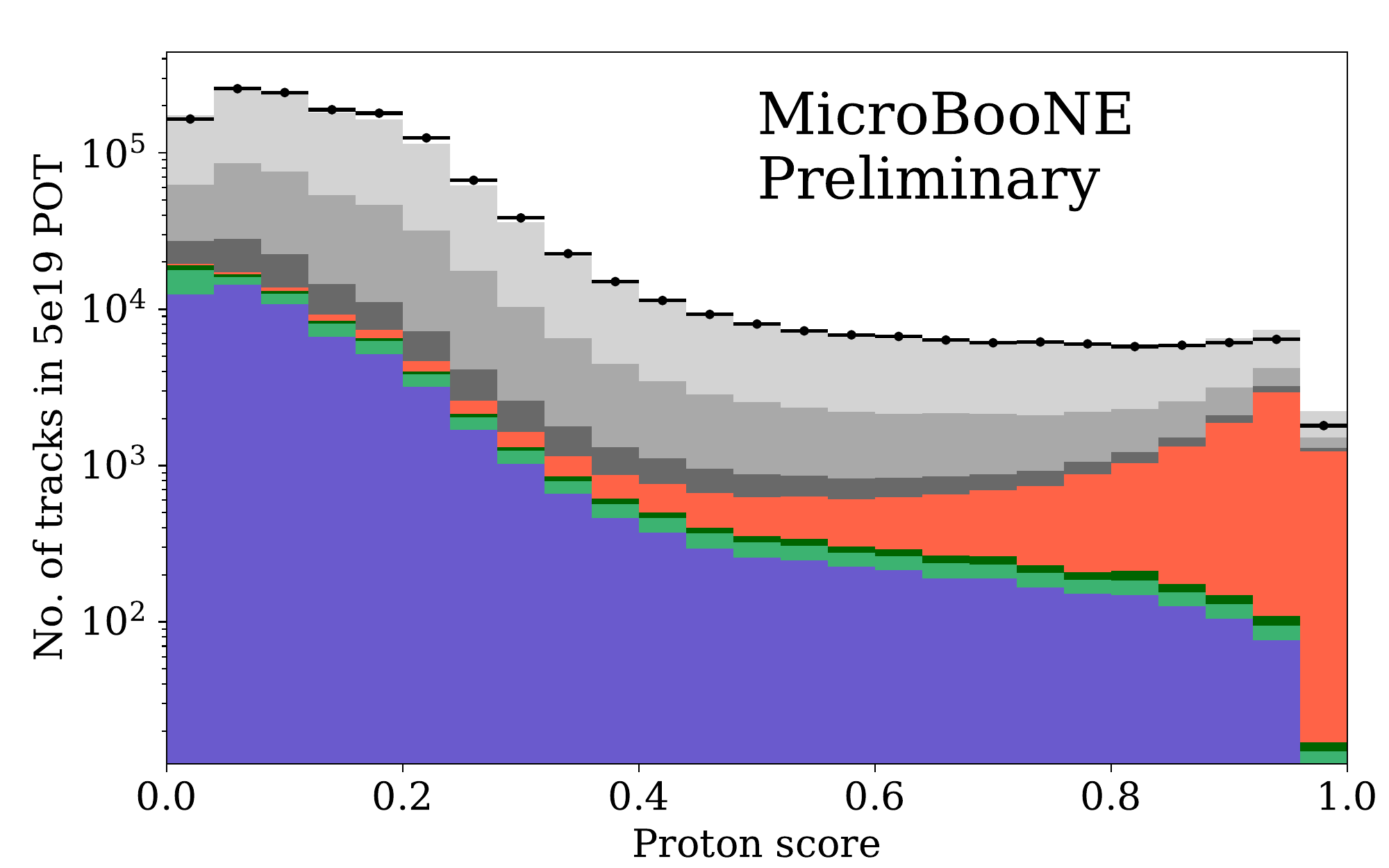}\hspace{2pc}%
  \end{minipage}
  \caption{Data to Monte Carlo comparisons of proton ID scores. The left plot
  shows linear scale, and the right plot shows the same thing in a log scale.
  The gray filled histograms are cosmic background tracks, the color filled
  histograms show neutrino-induced Monte Carlo tracks, and the black points
  show tracks in $5\times 10^{19}$ POT data. The dashed line shows the proton
  ID cut for the NC elastic pre-selection. Uncertainty bars are statistical
  only. 
  \label{fig:pscoredatamc}}
\end{figure}

Figure~\ref{fig:pscoredatamc} shows the decision tree proton score for all
tracks that are contained in the MicroBooNE detector for both a subset of the
neutrino beam data corresponding to $5\times 10^{19}$ protons on target (POT)
and the Monte Carlo simulation.

If we keep only tracks that have a proton score greater than 0.5 (anything more
likely than not to be a proton) we have an overall proton identification
efficiency of approximately 80\%. This does not include the reconstruction
efficiency for protons.

\begin{figure}[!htb]
  \centering
  \begin{minipage}[c]{16pc}
    \includegraphics[scale=0.46]{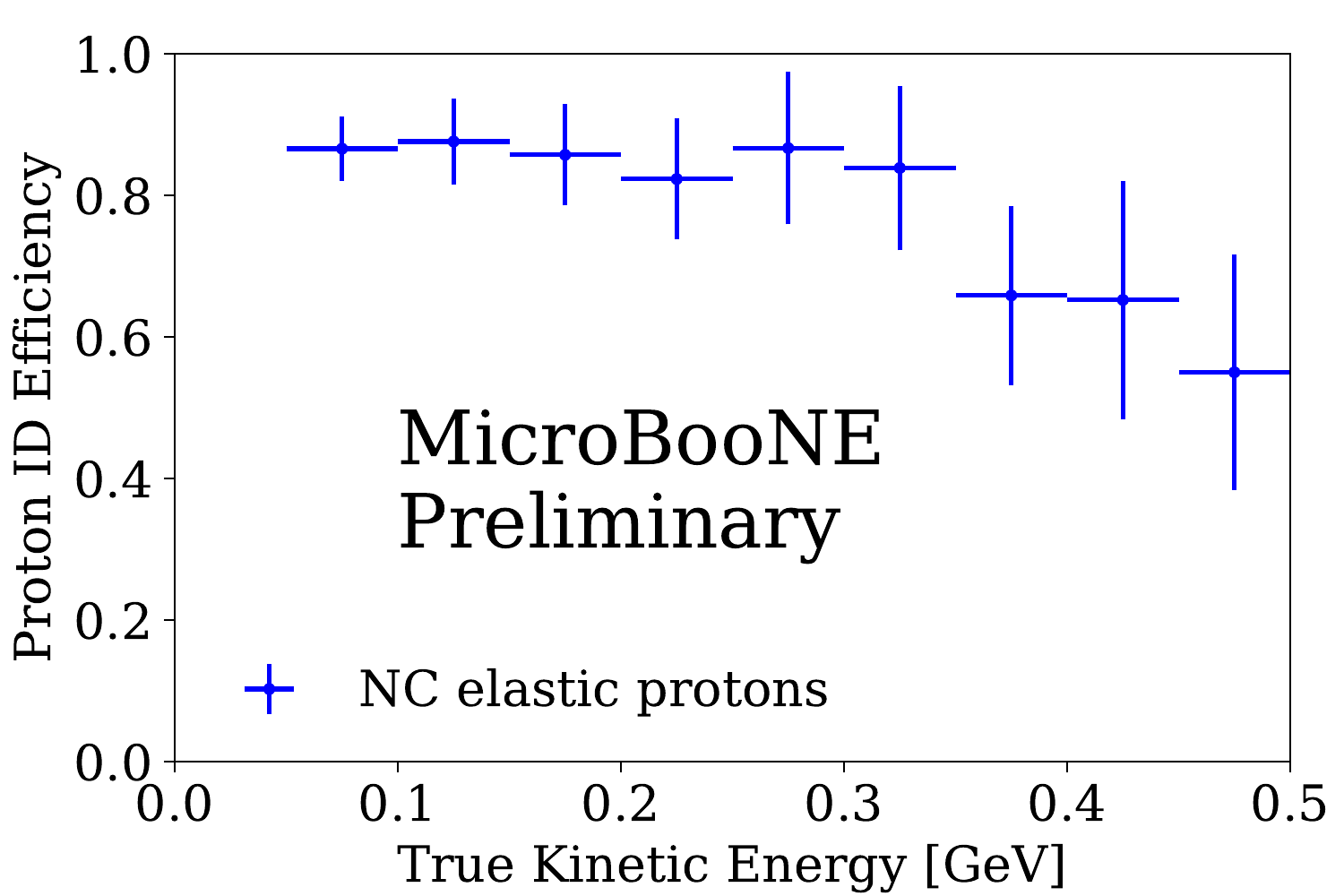}\hspace{2pc}%
  \end{minipage}\hspace{2pc}%
  \begin{minipage}[c]{16pc}
    \includegraphics[scale=0.46]{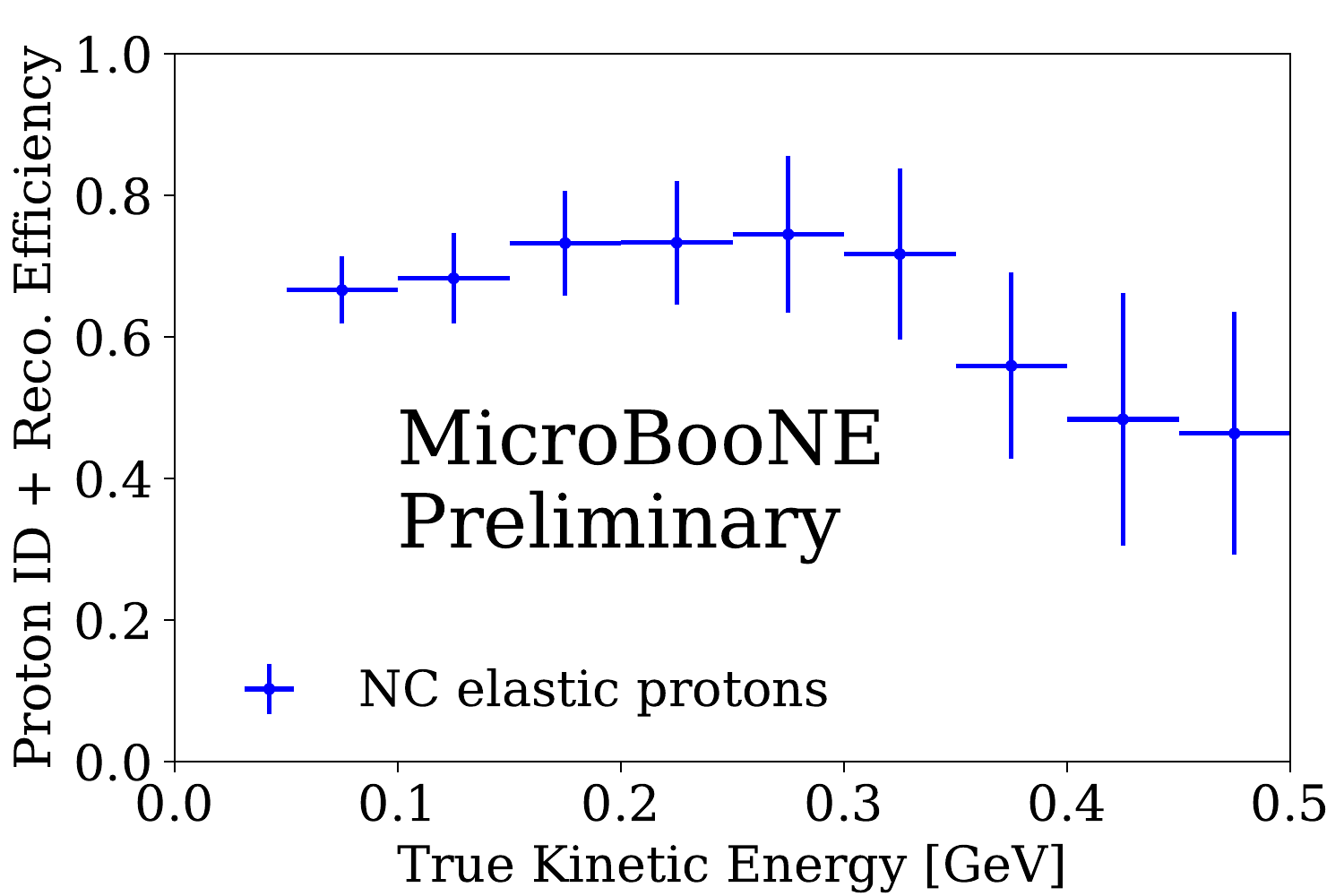}\hspace{2pc}%
  \end{minipage}
  \caption{The proton efficiency in neutral current elastic interactions. The
  left plot shows the NC elastic proton identification efficiency as a function
  of true proton kinetic energy. The proton identification efficiency is
  defined as the fraction of reconstructed true proton tracks that are
  correctly identified as protons. The right plots shows the overall NC elastic
  proton efficiency as a function of true proton kinetic energy. The overall
  proton efficiency is defined at the fraction of simulated NC elastic protons
  that are reconstructed and correctly identified at protons. The decrease in
  efficiency at larger proton kinetic energy is due to the increased
  probability of the proton reinteracting in the argon. The uncertainty is
  statistical only.
  \label{fig:effkence}}
\end{figure}

Figure~\ref{fig:effkence} shows the efficiency of protons from simulated
neutral current elastic events with a proton score greater than 0.5. Both the
proton identification efficiency and the overall, reconstruction plus
identification, efficiency as a function of true proton kinetic energy are
shown for NC elastic events. The average proton ID efficiency over all energies
is 75\% for NC elastic proton interactions. When including the efficiency of
track reconstruction, the average overall proton efficiency for NC elastic
events is 60\%. The proton kinetic energy range of interest for this analysis
is from 0.05 to 0.5 GeV, which end at the very beginning of the drop in proton
efficiency. The average proton ID efficiency in the $0.05-0.5$~GeV kinetic
energy range is 80\% for NC elastic protons and 64\% when including the
efficiency of track reconstruction.

\section{Selection}
\label{sec:selection}

We start the selection by making pre-cuts to remove a large amount of the
background. The remaining events are evaluated using a logistic regression
model to determine how signal-like they are (described in
Sec.~\ref{sec:logreg}). The logistic regression output gives us a parameter to
tune to maximize our ability to extract the strange axial form factor
parameters.

\subsection{Pre-Selection}\label{sec:presel}

  To reduce the cosmic background, we rejected any events that did not have a
  reconstructed flash with at least 6.5 photoelectrons within the 1.6~$\mu s$
  beam timing window. We select reconstructed tracks that were fully contained
  in the TPC fiducial volume and at least 2.5 cm long as potential proton
  candidates. The track is selected if it is classified as a proton with a
  greater than 50\% probability by the gradient-boosted decision tree
  classifier described in~\ref{sec:protonid}.

  \subsection{Logistic Regression of Selection Variables}
  \label{sec:logreg}

    Our NC elastic events of interest appear as a single isolated proton that
  produces scintillation light during the beam-spill window. The reconstructed
  variables that we use to select these events are: (1) the decision tree proton
  ID score, (2) the distance from the track to the reconstructed flash in the z
  direction, (3) the distance from the track to the reconstructed flash in the y
  direction, (4) the distance between the track candidate and the next closest
  reconstructed track, (5) whether or not the track is in the beam direction, (6)
  the distance between any reconstructed tracks IDed as neutrino-induced muons
  and the reconstructed beam flash, and (7) the distance between any
  reconstructed tracks IDed as neutrino-induced pions and the reconstructed beam
  flash.

  The proton ID score is the same as used in the pre-selection and described in
  detail in Sec.~\ref{sec:protonid}. The muon and pion background candidates
  are defined as any reconstructed track that has a muon ID or pion ID score
  greater than 0.5 from the boosted decision tree classifier. We define the
  one-dimensional distances between reconstructed tracks and reconstructed
  flashes as the difference in position between the midpoint of the track
  (defined by its endpoints) and the photoelectron-weighted center of the
  flash.

  To determine which events are NC elastic like based on these seven variables,
  we use them as input to a logistic regression model~\cite{Hosmer2005}. In
  logistic regression we fit a multi-dimensional sigmoid function to the signal
  and background data. The output is a score that can be used to determine how
  signal-like a data point is.
  \begin{equation*}
    S(g({\bf x})) = \frac{e^{g({\bf x})}}{1 + e^{g({\bf x})}} \,,
  \end{equation*}
  where $g({\bf x})$ is a linear combination of the selection variables, ${\bf x}$,
  \begin{equation*}
    g({\bf x}) = w_0 + w_1 x_1 + w_2 x_2 + ... + w_7 x_7 \,.
  \end{equation*}
  Here $x_1$ is the proton ID score, $x_2$ is the distance to the flash, etc.
  The set of weights, $w_0,...,w_7$ are determined from a fit to the data. We
  determined these weights using the StatsModels module~\cite{pystats} in
  Python to fit the model to a purely Monte Carlo data set.

\begin{figure}[!htb]
    \begin{minipage}[b]{20pc}
      \includegraphics[scale=0.44]{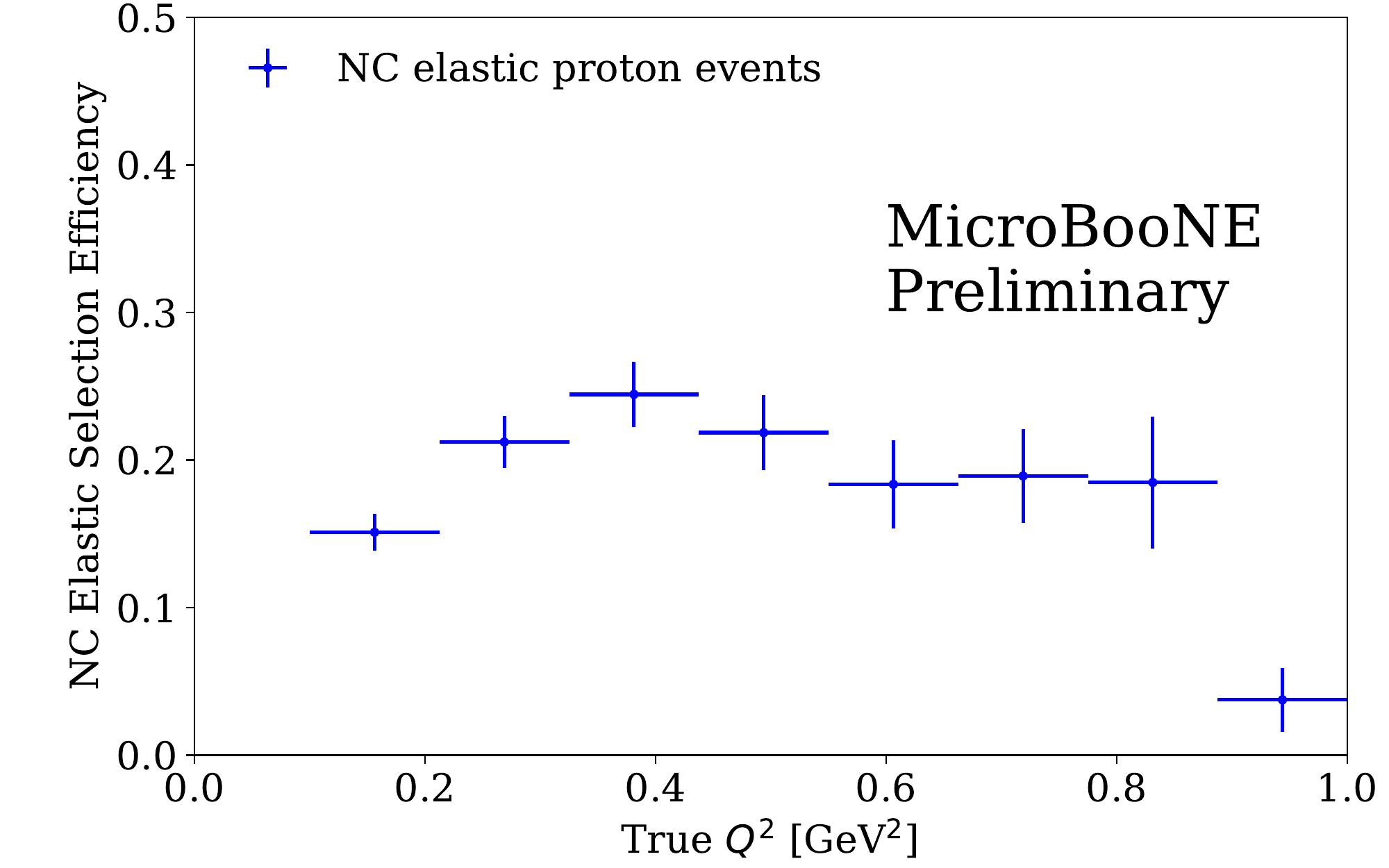}
    \end{minipage}\hspace{2pc}%
    \begin{minipage}[b]{14pc}%
    \caption{Neutral current elastic selection efficiency as a function of true
    four-momentum transfer for a logistic regression score cut of 0.8. The
    uncertainties are statistical only.
    \label{fig:efficiencyLR}}
    \end{minipage}
  \end{figure}

  If we choose a score cut of 0.8, for example, we get an overall signal
  efficiency of 19\% and purity of 18\%. Figure~\ref{fig:efficiencyLR} shows
  the signal selection efficiency as a function of true negative four-momentum
  transfer for a 0.8 score cut. The reconstructed four-momentum transfer is
  determined entirely from the proton kinetic energy using
  \begin{align*}
    Q^2_p &= -q^2 = -(\bf{p'}_p - \bf{p}_p)^2 \\
          &= -(E'_p - E_p)^2 + (\bar{p}'_p - \bar{p}_p)^2 \\
          &= 2 T_p M_p,
  \end{align*}
  where $\bf{p}$ is four-momentum, $E$ is energy, $\bar{p}$ is three-momentum,
  $M$ is mass, $T$ is kinetic energy determined by the length of the track, the
  $p$ subscript represents the proton in the neutrino-proton interaction, the
  prime represents the final state, and the proton momentum in the nucleus is
  assumed to be small compared to the final proton momentum. When the final
  proton momentum is close to the initial proton momentum in low $Q^2$
  interactions, this assumption isn't as strong and we rely more heavily on the
  nuclear model in the GENIE simulation.

\begin{figure}[!htb]
    \begin{minipage}[b]{20pc}
      \includegraphics[scale=0.35]{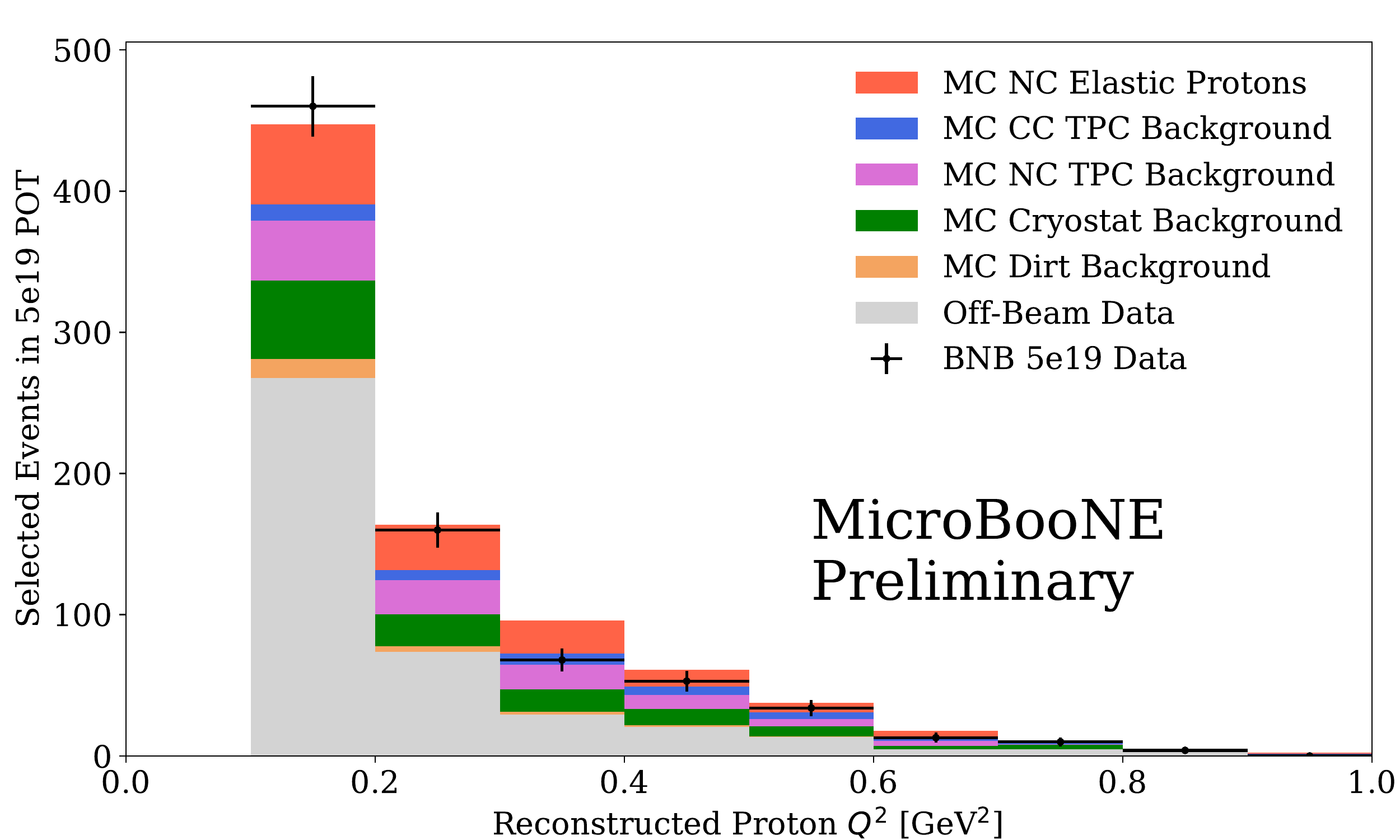} \\
      \includegraphics[scale=0.35]{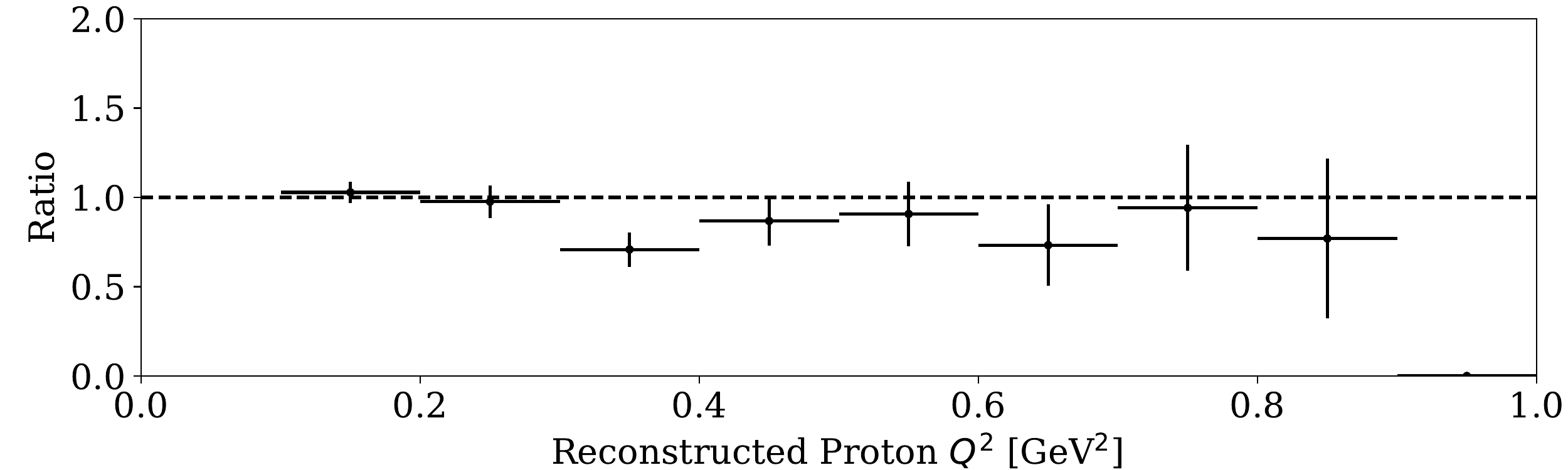}
    \end{minipage}\hspace{2pc}%
    \begin{minipage}[b]{14pc}
      \caption{NC elastic event selection given a logistic regression score cut
      of 0.8 as a function of reconstructed negative four-momentum transfer.
      The top plot shows the signal and itemized backgrounds compared to the
      BNB $5\times 10^{19}$ POT data in black. The bottom plot shows the ratio
      of BNB $5\times 10^{19}$ POT data to BNB MC with cosmic overlay and
      off-beam data. In both plots the uncertainties are statistical only. 
      \label{fig:recoq2LR}}
    \end{minipage}
  \end{figure}

\begin{figure}[!htb]
    \begin{minipage}[b]{20pc}
      \includegraphics[scale=0.35]{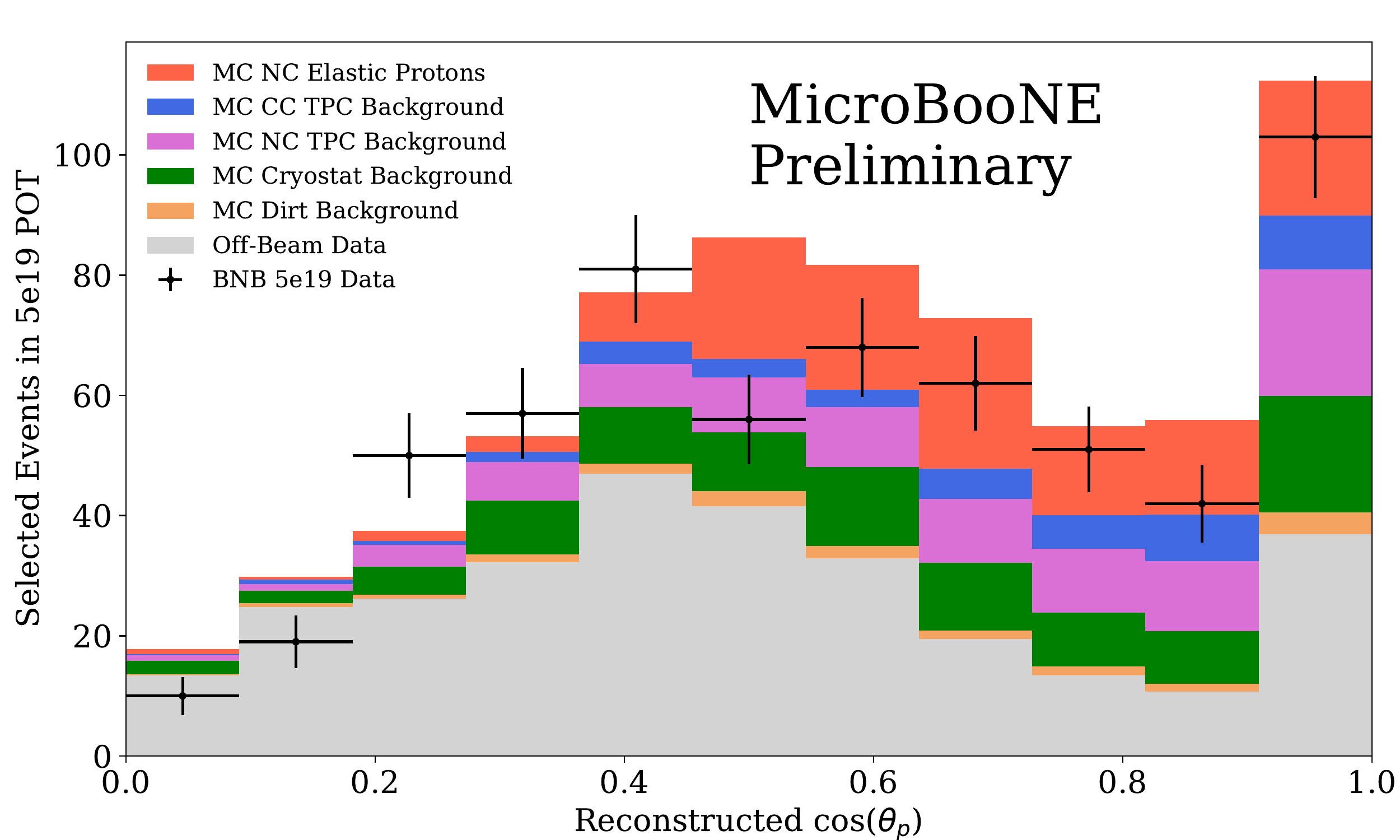} \\
      \includegraphics[scale=0.35]{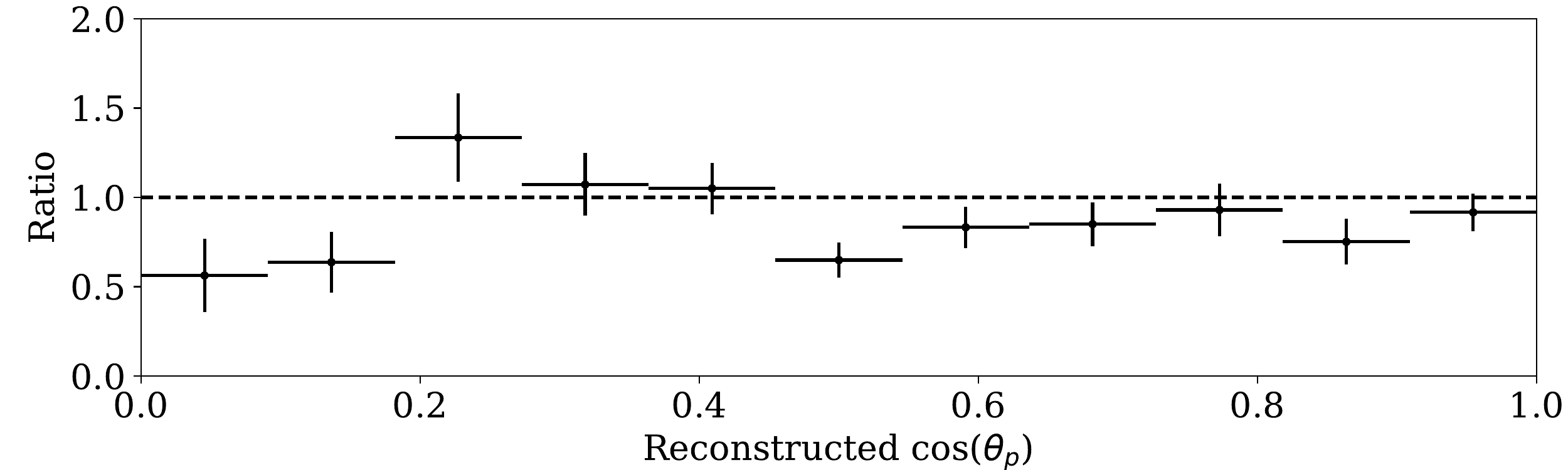}
    \end{minipage}\hspace{2pc}%
    \begin{minipage}[b]{14pc}
      \caption{NC elastic event selection given a logistic regression score cut
      of 0.8 as a function of reconstructed proton cos($\theta_p$), which is
      the proton angle with respect to the neutrino beam direction. The top
      plot shows the signal and itemized backgrounds compared to the BNB
      $5\times 10^{19}$ POT data in black. The bottom plot shows the ratio of
      BNB $5\times 10^{19}$ POT data to BNB MC with cosmic overlay and off-beam
      data. In both plots the uncertainties are statistical only.
      \label{fig:costhetaLR}}
    \end{minipage}
  \end{figure}

\begin{figure}[!htb]
    \begin{minipage}[b]{20pc}
      \includegraphics[scale=0.35]{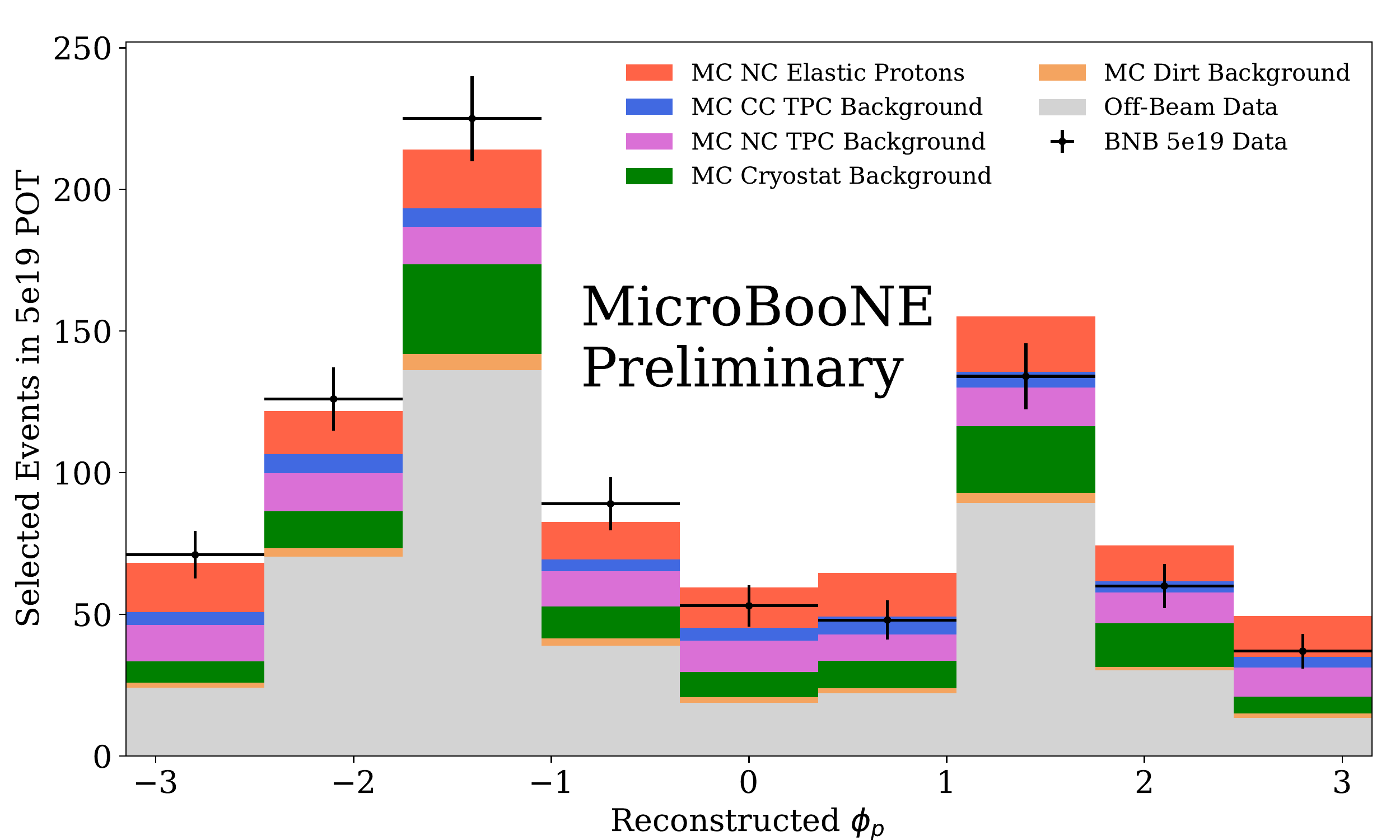} \\
      \includegraphics[scale=0.35]{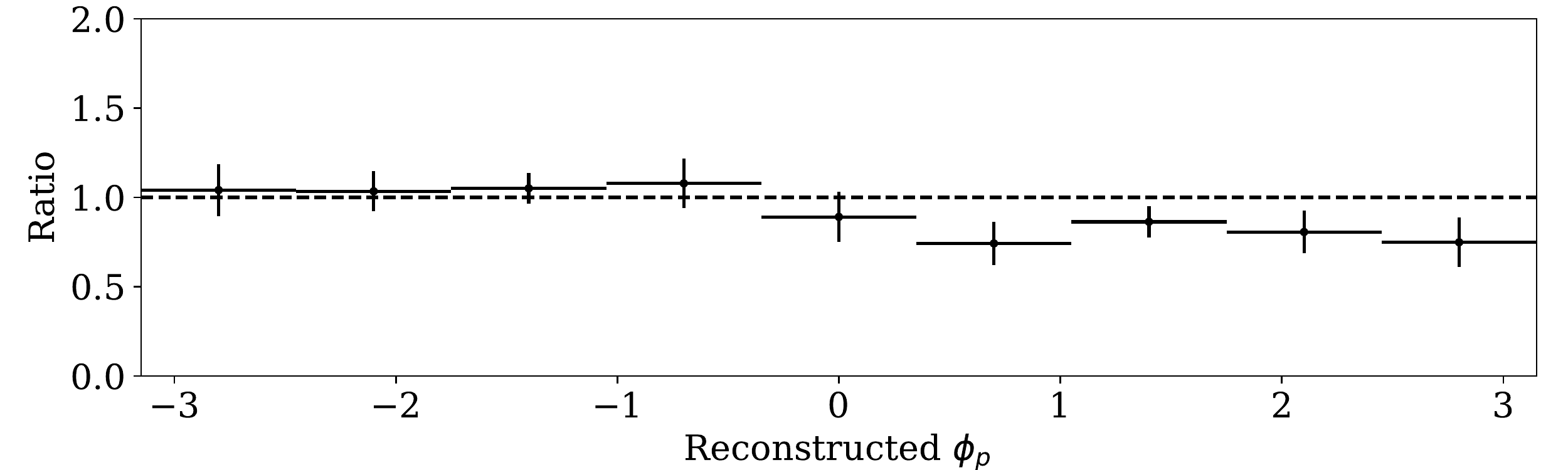}
    \end{minipage}\hspace{2pc}%
    \begin{minipage}[b]{14pc}
      \caption{NC elastic event selection given a logistic regression score cut
      of 0.8 as a function of reconstructed proton $\phi$. The top plot shows
      the signal and itemized backgrounds compared to the BNB $5\times 10^{19}$
      POT data in black. The bottom plot shows the ratio of BNB $5\times
      10^{19}$ POT data to BNB MC with cosmic overlay and off-beam data. In
      both plots the uncertainties are statistical only.
      \label{fig:phiLR}}
    \end{minipage}
  \end{figure}

  Figures~\ref{fig:recoq2LR},~\ref{fig:costhetaLR},~and~\ref{fig:phiLR} show
  the signal and backgrounds selected compared to data for a 0.8 logistic
  regression score cut. The number of simulated events and off-beam data events
  have been normalized to the $5\times 10^{19}$~POT of on-beam data.

\subsection{Remaining Backgrounds}

  The remaining backgrounds fall into four separate categories: (1) neutral
  current back interactions in the TPC, (2) charged current interactions in the
  TPC, (3) neutrino interactions outside of the TPC, and (4) cosmic interactions
  in time with the beam.

  The first category, NC backgrounds in the TPC, include NC elastic
  interactions with a {\it neutron} (35\%), NC elastic interactions with a
  correlated neutron-proton pair (35\%), NC resonant interactions (25\%), and NC
  DIS interactions (4\%). The last two, resonant and DIS, are mainly due to
  tracks being reconstructed poorly or not reconstructed at all. The NC
  interaction backgrounds are due to the fact that neutrons are very difficult to
  detect in LArTPCs. In the case of the NC elastic {\it neutron} interactions, we
  select events in which the neutron scatters with a proton mimicking a NC
  elastic neutrino-proton event. In the case of NC interactions with correlated
  neutron-proton pairs, we select the proton from the initial interaction, but
  the neutron goes undetected. 

  The second category, CC interactions in the TPC, are almost entirely from the
  muon track being mis-reconstructed, mis-identified as a cosmic, or not
  reconstructed at all. In these events, a secondary proton from the
  interaction or a coincident cosmic proton can be selected as a
  neutrino-induced proton. 

  The third category, TPC-external interactions, occur when a neutrino from the
  beam interacts outside of the active TPC volume. A neutron from this
  interaction can enter the TPC and elastically scatter with a proton. The
  neutrino interaction occurs in the liquid argon inside of the cryostat
  $\sim$70\% of the time, and outside of the cryostat in the dirt upstream of
  the detector hall the other $\sim$30\% of the time.  These cryostat-external,
  ``dirt", interactions are simulated separately from the interactions inside
  the cryostat. The normalization of this dirt sample was determined by fitting
  the flash position in the beam direction to data. There is an observed excess
  of flashes from dirt events in the upstream end of the detector. This region
  is excluded in the analysis with fiducial cuts.

  The last category, cosmics in time with the beam, is also the largest.
  However, we can determine the rate and distribution of this background by 
  taking data while the neutrino beam is off. This is the
  ``off-beam" data set. These events can then be subtracted from the ``on-beam"
  data.

\section{Conclusions}
\label{sec:summary}

  The contribution of the strange quark spin to the spin of the proton is an
  open and interesting question. A lot can be learned about this physics
  through the strange part of the axial form factor, including the net strange
  spin contribution, $\Delta s$, which is simply the value of the strange axial
  form factor, $G_A^s$, at zero four-momentum transfer. MicroBooNE's ability to
  detect low-energy protons translates into an ability to measure $G_A^s$ at
  low four-momentum transfer. 

  Selecting isolated proton-like reconstructed tracks near a beam flash gives
  us sets of events with an enhanced fraction of NC elastic events. The
  efficiency and purity of this selection, as well as the agreement between the
  selection in Monte Carlo and the selection in the BNB $5\times 10^{19}$ POT
  data set can be tuned to optimize the sensitivity to $\Delta s$ using the
  output of our logistic regression model.

  Assuming the current level of neutral current elastic proton event selection
  efficiency, we expect to select on the order of 1000 NC elastic events in the
  full MicroBooNE data set with a $Q^2$ down to 0.1 GeV$^2$. The next stage of
  this analysis will include updated and improved energy calibration which is
  expected to increase the agreement between data and simulation. A full
  treatment of the systematic uncertainty due to the physics models, the proton
  selection, and the detector efficiency will also be included.  This should
  allow us to extract the strange axial form factor parameters with a greater
  precision than has previously been possible in neutrino-nucleon scattering
  experiments.


\end{document}